# Origin of shallow n-type doping in AlN and Al-rich AlGaN

Yujie Liu,[1] Sieun Chae,[1,2] and Emmanouil Kioupakis[1,a]

[1]*Department of Materials Science and Engineering, University of Michigan, Ann Arbor, Michigan 48109, USA*

[2]*School of Electrical Engineering and Computer Science, Oregon State University, Corvallis, Oregon 97331, USA*

**Achieving efficient n-type doping in AlN, a representative ultrawide bandgap (UWBG) semiconductor, remains a longstanding challenge that limits its application in high-power electronics and deep-ultraviolet optoelectronics. Conventional dopants in AlN often introduce deep levels or form compensating complexes, leading to low free-carrier concentrations. In this work, we combine first-principles defect calculations with a structural search method tailored to explore metastable configurations to systematically investigate donor-type defects in AlN. Our results reveal that the aluminum interstitial ($Al_i$) can exhibit shallow-donor behavior in specific metastable configurations that were previously overlooked. This discovery expands the understanding of n-type dopability in AlN, and highlights the critical role of metastable defects in modulating electronic properties.**

Aluminum nitride (AlN), an ultrawide band gap (UWBG) semiconductor with a direct band gap of ~6.2 eV, offers exceptional properties such as high breakdown electric fields, superior thermal conductivity, and intrinsic radiation hardness.[1] These attributes make AlN highly attractive for power electronics, ultraviolet optoelectronics, and emerging quantum technologies. Notably, AlN is widely used as a template and buffer layer in deep ultraviolet (DUV) light-emitting diodes (LEDs),[2] in which its lattice compatibility and UWBG enable efficient emission at wavelengths shorter than 280 nm.[3] Despite this technological promise, achieving efficient n-type doping in AlN remains a fundamental challenge. One of the key limiting factors is the formation of deep donor-related defects, commonly known as DX centers, which are deep-level defect configurations that trap electrons and severely limit the carrier concentration.[4] The concept of DX centers was originally introduced to explain the anomalous behavior of Si donors in AlGaAs,[5] and these insights have been extended to UWBG semiconductors, in which strong ionicity and poor dielectric screening enhance the likelihood of DX formation, particularly in nitrides such as AlN.[6]

---

[a] Electronic mail: kioup@umich.edu

Understanding the behavior of such defect states is thus essential to improving dopability and enabling the full potential of AlN in optoelectronic applications.

Experimental studies have previously uncovered evidence of deep DX-center formations both for Si[7–23] and for Ge[24,25] donors in AlN and Al-rich AlGaN, with the underlying mechanism explained as self-compensation processes. This is based on observations of significantly decreased carrier concentration and electrical conductivity due to higher dopant activation energies. However, a wealth of recent works revealed clues for shallow Si[22,26–33] and Ge[34–36] donors in AlN, both via optical identification such as cathodoluminescence (CL) measurements by Matthews *et al.*[31] and Hauwiller *et al.*[32], photoluminescence (PL) measurements by Mondal *et al.*[33], and exciton emission spectra by Neuschl *et al.*[28], as well as the direct observation of higher carrier concentration and higher n-type conductivity by Breckenridge *et al.*[27] and Almogbel *et al.*[37] Specifically, low ionization energies of ~75 meV were demonstrated via an non-equilibrium process based on ion implantation of Si and Ge, followed by annealing under above-gap illumination.[27,29,35] The reported experimental values for these shallow and deep Si incorporations in AlN and Al-rich AlGaN are summarized in Table SI. Moreover, previous works have also reported the formation of cation vacancy-related second-nearest-neighbor $V_{III}$-$n$•donor defect complexes (i.e., point-defect complexes of a group-III cation vacancy with $n$ substitutional donor impurities) in both Si- and Ge-doped AlN and AlGaN samples via room-temperature PL spectroscopy.[20,25,27,34,38] These studies concluded that impeding the formation of these defect complexes can greatly improve the optical properties of these samples, and identified them as a dominant compensation mechanism in the high Si-doping limit.[39]

On the other hand, computational works provide evidence for DX-center behavior both for Si and for Ge donors in wurtzite AlN and Al-rich AlGaN. This finding is consistent for different levels of exchange-correlation such as the local density approximation (LDA)[40–43] and the more accurate hybrid functional of Heyd, Scuseria, and Ernzerhof (HSE).[44,45] Despite the wealth of experimental evidence for shallow-donor formations in AlN, there is a lack of theoretical explanation about how shallow n-type doping can occur in AlN, especially in light of the established understanding of deep DX-center formation in Si- and Ge-doped AlN. Interestingly, Yan *et al.* have revealed the importance of Al vacancy ($V_{Al}$) with oxygen substitutions ($O_N$, which also behave as DX-centers in Al-rich AlGaN) forming defect complexes in AlN,[46,47] while Lyons *et al.* uncovered that carbon interstitials could contribute to n-type-conductivity.[48] These works further motivate the investigation of possible complexes between Si/Ge donors and other defects in AlN to uncover the origins of the shallow n-type dopability.

In this study, we apply first-principles calculations to uncover the mechanism of shallow n-type doping in Si- and Ge-implanted AlN, and by extension to Al-rich AlGaN. First, through a detailed structural and energetic analysis of $Si_{Al}$ DX centers in AlN, we find that the in-plane bond-disrupted configuration (denoted as α-BB by Park and Chadi[40], or $DX_2$ by Silvestri *et al.*[45]) is more stable than the c-axis bond disruption (denoted as γ-BB or $DX_1$) for $Si_{Al}^-$. Next, we uncover that the shallow Si donor state is metastable, separated by energy barriers of ~360 meV for $DX_1$ and ~750 meV for $DX_2$ from the two distinct deep DX configurations. At the same time, we find that there is no kinetic barrier for shallow Ge donors to transition to the DX geometry, which are therefore unstable in the shallow geometry. Subsequently, we perform a detailed investigation of various defects and defect complexes in AlN, confirming the shallow donor characteristics of metastable Al interstitials ($Al_i$). In particular, we discuss the formation of DX centers and electron lone pairs for different charge states of $Al_i$. Our findings provide critical insights into the n-type dopability of AlN, underline the importance of controlling defect behavior to achieve highly conducting n-type AlN, and offer guidance on enabling next-generation high-power electronics and deep-ultraviolet optoelectronic applications.

# Results

**DX centers for $Si_{Al}^-$ and $Ge_{Al}^-$**

We first revisit the $Si_{Al}^-$ and $Ge_{Al}^-$ DX-center configuration, formation, and ionization by computing the equilibrium atomic coordinates and energies utilizing the ShakeNBreak (SNB) code[49,50] to determine the global and local minima.[51] The corresponding formation energy results are shown in Fig. 1 (a) and (c). In contrast to previous computational results by Gordon *et al.*[44] and Silvestri *et al.*[45], while in agreement with Gaddy[52], we find that the lowest-energy DX-center configuration for $Si_{Al}^-$ is the α-BB/DX$_2$ geometry, as shown in the rightmost inset atomic structure of Fig. 1 (b), instead of the γ-BB/DX$_1$ geometry with an energy higher by ~130 meV. See Figure S1 and Table SII in supplementary materials for more details about the two types of DX centers for $Si_{Al}^-$. For $Ge_{Al}^-$, we found the DX$_1$ configuration to be the only stable state, consistent with previous calculations.[44,53] Moreover, the transition level $\epsilon(+/-)$ is ~280 meV for $Si_{Al}$ and ~980 meV for $Ge_{Al}$, indicating that both Si and Ge DX centers are deep donors.

We next examine the possible metastability of shallow $Si_{Al}$ and $Ge_{Al}$ donors in AlN. To accomplish this, we generate atomic coordinates of Si and Ge incorporation in the unperturbed, shallow geometry and apply climbing image nudged elastic band (cNEB) calculations[54] to determine the kinetic barrier to the deep DX atomic configurations. As shown in Fig. 1 (b), shallow Si donors exhibit a migration barrier of ~360 meV to the α-BB/DX$_2$ state, and ~780 meV to the γ-BB/DX$_1$ state (see Figure S2 in supplementary materials). To estimate the rate for defect hopping, we use the Arrhenius equation $R = \mu \exp\left(\frac{-E_a}{kT}\right)$,[55] where μ is the attempt frequency. At room temperature, with an activation energy (migration barrier) of $E_a = 360$ meV for conversion of shallow donors to DX$_2$ centers (the more likely to occur), and an attempt frequency μ of 10 THz (i.e., of the order of magnitude of TO phonon frequencies in AlN), the defect hopping rate R is estimated to be ~$10^6$ Hz, suggesting that the conversion of metastable shallow Si donors to stable DX centers is too fast even at room temperature, and cannot be suppressed effectively. Therefore, the formation of metastable shallow Si is not the reason for the shallow doping observed in experiments. This is further corroborated by the lack of an energy barrier for the conversion of metastable unperturbed substitutional Ge donors to deep DX centers, Fig. 1 (d), which contradicts the experimental evidence for shallow donors in Ge-implanted AlN. This evidence suggests that a mechanism other than metastable Si or Ge donors is responsible for the experimentally reported shallow n-type doping in AlN.

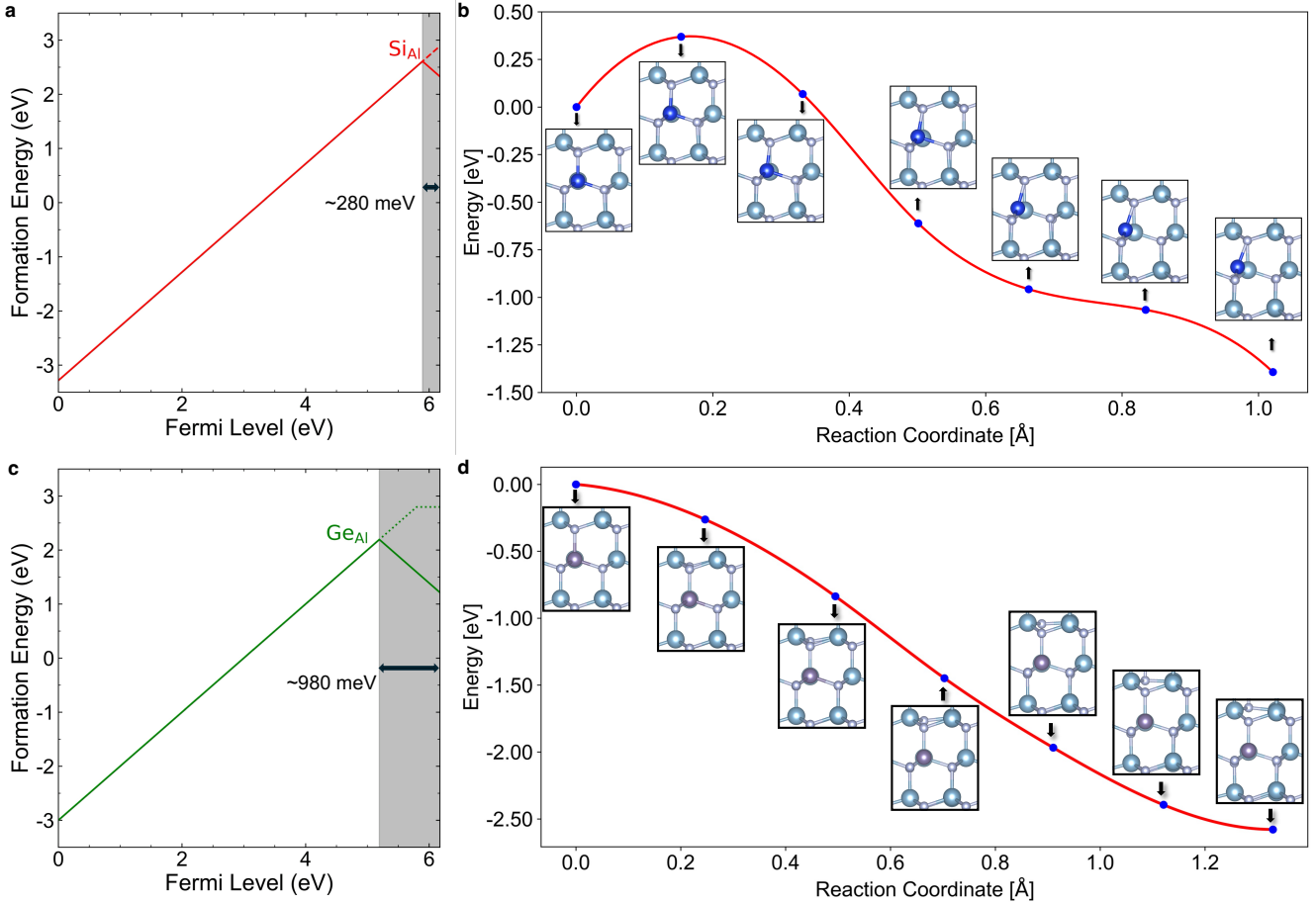

FIG. 1. (a) Formation energy plot of $Si_{Al}$ and (c) $Ge_{Al}$ using hybrid functionals. The transition level $\epsilon(+/-)$ is denoted. (b) cNEB calculations for the transition from the shallow donor geometry (left) to the stable deep DX center (right) of Si α-BB/DX$_2$ ($Si_{Al}^-$), and (d) for Ge donors ($Ge_{Al}^-$) using different geometries as interpolations in AlN. Both curves are smoothed through spline fitting. The finite-size corrections for the energies of different images treated as charged defects are included. The low and no energy barriers for shallow-to-deep transition processes of $Si_{Al}^{-1}$ and $Ge_{Al}^{-1}$ indicates that metastable shallow Si and Ge donors are unlikely to occur.

**Thermodynamics of defects in AlN**

In light of the previous results that preclude the existence of metastable shallow Si and Ge donors, we explore alternative defects and defect complexes as the source of the experimentally observed shallow dopants in AlN. These include Si- and Ge-related complexes with other native point defects such as vacancies or interstitials, as well as high-energy native point defects that form through the process of Si/Ge implantation and annealing. Our calculated results for the formation energy of these alternative defects in their various charge states are shown in Fig. 2 (a). Our results suggest that Al interstitials ($Al_i$) could provide an alternative mechanism for the origin of the shallow doping. However, the most stable configurations for $Al_i$, depicted in Figs. 2 (e) and (f), behave as DX-centers with a transition level $\epsilon(+1/-1)$ of ~400 meV below the CBM, i.e., slightly deeper than but comparable to the $Si_{Al}$ DX center. Therefore, the thermodynamically stable $Al_i$ interstitial configurations do not behave as shallow donors. Moreover, we find that Al interstitials have a very high formation energy (more than 10 eV) under n-type conditions and are unlikely to form in thermodynamic equilibrium. However, it is likely that such metastable defects can be generated and kinetically stabilized through out-of-equilibrium experimental processes that include ion implantation[27,29,35], vapor phase-involved growth at high temperature[28,30,34], and epitaxial growth under high Al pressure at low temperature[31] that have been applied in various experiments. On the other hand, none of the defect complexes we examined between Si/Ge and vacancies or interstitial defects are found to exhibit shallow-donor behavior. We note that previous works reported similar calculations for the intrinsic defects in AlN, including $Al_i$. However, certain limitations such as the accuracy of the exchange-correlation functional[56–59] or the lack of consideration of the $Al_i^{-1}$ charge state[60,61] precluded conclusions regarding its behavior as a donor.

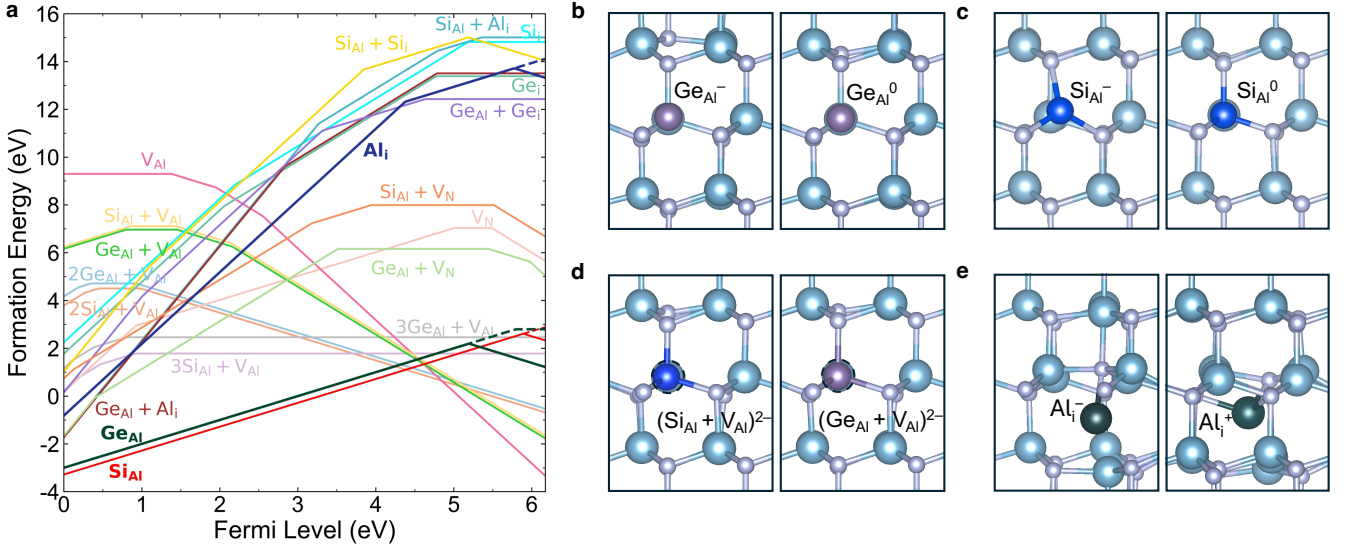

FIG. 2. (a) Formation energy as a function of Fermi level for intrinsic vacancies ($V_N$, $V_{Al}$), cation interstitials ($Ge_i$, $Si_i$, $Al_i$), cation substitutions ($Si_{Al}$, $Ge_{Al}$), and relevant defect complexes for donors in AlN under N-rich conditions. (b) - (e): Atomic configurations of $Ge_{Al}$, $Si_{Al}$, $nSi_{Al}/Ge_{Al}$ + $V_{Al}$ (for $n$ up to 3), and $Al_i$ in AlN. The bold solid red and dark green lines, $Si_{Al}$ and $Ge_{Al}$, adopt the $DX_1$ configurations with an extra electron, while $Si_{Al}^0$, $Si_{Al}^+$, and $Ge_{Al}^+$ adopt the shallow geometry in the region of dashed red and dark green lines. The compensation effect of $Al_i$ shallow donors (solid bold darked blue line) by $V_{Al}^{3-}$ (pink line) is suppressed through the formation of $V_{Al}$ complexes with substitutional Si and Ge, shown as $Si_{Al} + V_{Al}$ (gold line) and $Ge_{Al} + V_{Al}$ (lime green line), which have been observed in PL peaks by Washiyama *et al.*[25] and adopt the shallow geometry since the Si/Ge-N bond is not disputed in (d). In (e), the most stable $Al_i^-$ (denoted as dark cyan atoms) states represent a dumbbell-like DX-center geometry, and the $Al_i^+$ adopt the γ-BB like geometry.

**Metastable $Al_i$ as metastable shallow donors**

To further analyze the properties of Al interstitials as possible intrinsic shallow donor defects, we explore the energies and local atomic configurations of the various stable and metastable octahedral-site $Al_i$ geometries in the +1, 0, and –1 charge states. Al interstitials provide an excess of three unbonded electrons that, like the free electrons provided by substitutional Si/Ge DX centers, can induce the breaking of nearby Al-N bonds and result in similarly distorted DX center configurations. However, the three additional electrons provided by Al interstitials can induce more than a single BB compared with the one excess electron of substitutional Si/Ge DX centers, enabling a broader array of possible atomic configurations. To explore the landscape of possible stable and metastable $Al_i$-related atomic configurations, we again apply bond distortions for $Al_i$ in three charge states with the SNB code. The energies as a function of bond distortion factor $k$ of the nearest neighbors (see Methods for the definition of $k$) for stable and metastable $Al_i^{+1}$, $Al_i^0$ and $Al_i^{-1}$ are shown in Figs. 3 (a), (b), and (c), along with the local atomic coordinates and the charge distribution of the mid-gap $Al_i$ state. In terms of the energetics, our results indicate that $Al_i^{+1}$ and $Al_i^0$ do not exhibit metastable states, since the total energy difference near $k = 0.3$ are within 10 meV, and the local geometry near $Al_i^{+1}$ and the distorted defect-neighbor distances $d$ of these three states are nearly identical. On the other hand, three metastable $Al_i^{-1}$ states are predicted at $k = -0.2$ for #1, #3 showing the same total energy and atomic structure at $k = -0.3, -0.1, 0$, and #2 at $k = 0.2$. The most stable $Al_i^{-1}$ geometry and the associated charge density corresponding to the occupied midgap states, is illustrated in Fig. 3 (c). Meanwhile, other observed metastable $Al_i^{-1}$ states have total energy differences of ~600 meV (#1), ~400 meV (#2), and ~300 meV (#3) higher than the ground state, respectively. These metastable $Al_i$ states are shallower with ionization energies of ~71 meV (#1), ~171 meV (#2), and ~202 meV (#3), as depicted in Fig. 4 (a).

To estimate the concentrations of metastable $Al_i^{-1}$, we apply the Boltzmann distribution $n_i = n_0 exp(-\frac{\Delta E_i}{kT})$ to obtain the relative population of each $Al_i^{-1}$ state. At room temperature (~300 K), the population of all three metastable states are under 0.01%. Our calculated values from cNEB for the energy barrier (~850 meV) for metastable #3 $Al_i^{-1}$ to transit to either metastable #1 or #2 (see Figure S3 in supplementary materials), in combination with the Arrhenius equation, predicts a hopping rate of 1 s$^{-1}$, i.e., we predict an equilibration between different metastable $Al_i^{-1}$ states. However, at the experimental annealing temperature range of 1200 °C ~ 1500 °C,[25,27,29] the concentration of stable $Al_i^{-1}$ decreases to ~85 %, indicating a ~15% (1% for #1, 4% for #2, 9% for #3) combined concentration of all identified metastable $Al_i$ at elevated temperatures. On the other hand, since the stable-to-metastable $Al_i^{-1}$ barrier is low and can be overcome rapidly even at room temperature, these high-temperature concentrations may not persist once the sample is cooled to room temperature. Nevertheless, as electronic devices heat under operating conditions, we expect that a higher fraction of $Al_i$ will adopt the metastable configurations and thus increase the free-electron concentrations.

As for the structural properties, the local geometry of $Al_i$ in all charge states resembles a DX-center-like configuration with a nearby vertical or horizontal broken Al-N bond. For $Al_i^{+1}$, the most disrupted N atom adopts a near-planar hexagonal configuration [Fig. 3 (a)], with the three in-plane Al-N bonds having a length of 1.9 Å, similar to the Al-N bond length in pristine AlN. However, stable $Al_i^0$ adopt a deeper DX configuration, in which the most disrupted neighboring N atom shifts further upward [Fig. 3 (b)], forming a longer vertical Al-N bond of ~2 Å. Furthermore, stable $Al_i^{-1}$ possesses one longer newly-formed Al-N bond of ~2.11 Å (12.8% elongation) and one shorter bond with the vertical disrupted N atom of ~1.82 Å (4.2% contraction), while $Al_i^{+1}$ forms two typical Al-N bonds with the length of ~1.92 Å (~1%, negligible), and one shorter with similar bond length to the 4.2%-contracted one in $Al_i^{-1}$. See Figure S4 for details of relevant bond lengths of stable $Al_i^{-1}$ and $Al_i^{+1}$. Other than these, for all three metastable $Al_i^{-1}$, the neighboring N atoms exhibit planarized distortions with the Al interstitials, resembling the α-BB/$DX_2$-like configurations. Furthermore, in contrast to the $Si_{Al}$ DX center, where the mid-gap state typically arises from a broken Al–N bond state, our results reveal a distinct mechanism for metastable Al interstitials. The mid-gap states are formed by a combination of electron lone pair formation and Al–N bond-breaking distortions. This is clearly evidenced by the projected band structures and the charge-density isosurfaces shown in Fig. 3, where the occupied mid-gap states exhibit asymmetric, lone pair-like distributions centered on the $Al_i$ site, and the band projections indicate that these deep mid-gap states arise primarily from $Al_i$ orbitals, evidenced by its strong spectral weight. These states are flat and isolated from the conduction and valence bands, indicating strong spatial localization and the ability to trap charge carriers. The formation of electron lone pairs alters the local orbital overlaps in specific directions, further facilitating the breaking of a nearby Al-N bond and stabilizing the interstitial configuration. Consequently, both stable and metastable $Al_i$ induce bond distortions and deep mid-gap states, highlighting the rich structural and electronic properties associated with aliovalent interstitials in wide-gap nitrides that do not behave like conventional donors or shallow traps.

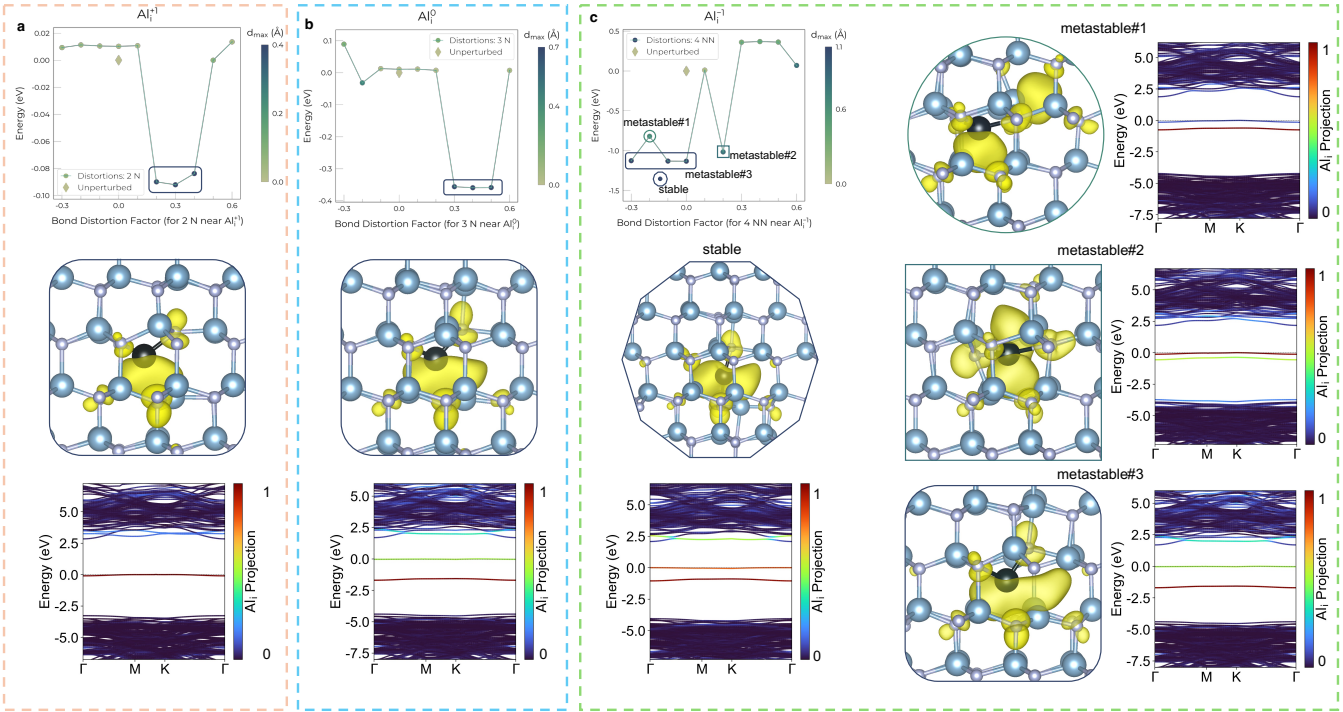

FIG. 3. Energies of stable and metastable geometries of (a) $Al_i^{+1}$, (b) $Al_i^0$, (c) $Al_i^{-1}$ as determined by the SNB method and code. The distorted defect-neighbor distances are denoted as $d_{max}$ and indicated by the side color bar. We also plot the local atomic geometry in the vicinity of the $Al_i$, as well as the charge density corresponding to the occupied midgap state(s) for the stable (rectangle and decagon denoted in (a) – (c)) and metastable (circle and square denoted in (c)) configurations in the top SNB plot. The interstitial Al atoms are specially highlighted by the black spheres in the lattice. The isosurfaces indicate the charge-density distribution of the midgap $Al_i$ states that accommodate the excess electrons, and are set to 5% of the maximum value. The bottom panel in each subplot presents the electronic band structures projected onto the $Al_i$ atom, where the color gradient represents the spectral weight of the $Al_i$. These projections reveal distinct deep midgap states corresponding to the defect levels formed by the symmetry-breaking distortions. For all three charge states, one of the neighboring Al/N atoms forms a DX-center-like configuration with a vertical or horizonal Al-N BB, similar to the $DX_1$ or $DX_2$ feature in $Si_{Al}$. For $Al_i^{-1}$ specifically, we identify one stable and three metastable configurations that are located ~600, ~400, and ~300 meV higher in energy. Compared to $Al_i^{+1}$, $Al_i^{-1}$ hosts an intrinsic electron lone pair due to the two extra electrons, as evidenced from the extended charge distribution for the stable state and two different metastable charge distributions in (c).

# Discussion

Through a comprehensive evaluation of possible defects and defect complexes in n-type doped AlN, we confirm the shallow-donor nature of metastable $Al_i$, which emerges as a plausible origin for the experimentally observed shallow donor behavior in ion-implanted samples. Importantly, this conclusion holds for both Si and Ge donors, as our defect formation energy calculations and charge-state transition levels indicate a similar trend for both donor species. This universality suggests that the stabilization of $Al_i$ is mediated by the formation of $V_{III}$–n•donor complexes, as demonstrated in prior experimental studies[20,62], which effectively immobilizes the cation vacancy and prevents its recombination with the displaced interstitial aluminum. As a result, $Al_i$ persists as an isolated and electrically active shallow donor. This mechanism appears intrinsic to the ion implantation process in AlN, rather than being specific to the chemical identity of the implanted donor. Therefore, controlling the interplay between complex formation and $Al_i$ diffusion becomes essential for optimizing the electrical activation and stability of doped AlN.

The presence of electron lone pairs in $Al_i$ can significantly affect the atomic mobility and diffusion properties of the system. Specifically, the lone pair electrons on the donor species introduce additional repulsion and localization effects, which elevate the diffusion barrier of the defect complex. As shown in Fig. 4 (b), the in-plane diffusion of $Al_i$ along the $a$-direction exhibits a relatively low barrier of ~0.5 eV, as it does not induce significant structural deformation. In contrast, diffusion along the $c$-direction shows a significantly higher barrier of ~3.0 eV, attributed to the structural distortions induced by the lone pair electrons. Diffusion in other directions would involve further bond-breaking processes, suggesting that the overall migration barrier of $Al_i$ remains high. As a result, $Al_i$ exhibits reduced mobility compared to typical isolated interstitials or other point defects. This reduction in mobility has important implications for the stability and distribution of defects during both synthesis and post-growth annealing processes.

Beyond the diffusion dynamics, our results reveal that the $V_{III}$–n•donor defect complex is not merely optically detrimental. Traditionally, the focus on defect complexes in wide-bandgap semiconductors such as AlN has revolved around their impact on optical and transport properties, particularly their role in acting as nonradiative recombination centers. However, our findings extend this perspective by demonstrating that the $V_{III}$-n•donor complex can serve as a source of electrically active species, particularly Al interstitials, through ion implantation. Moreover, the formation of $(Si_{Al}/Ge_{Al} + V_{Al})^{2-}$ can stabilize the Si/Ge donors in the +1 charge state, i.e., the shallow atomic configuration, as demonstrated in Fig. 2 (d), avoiding their incorporation as DX-centers that would suppress the carrier concentration. The Al interstitials act as shallow donors and contribute significantly to the n-type conductivity in doped AlN. Their simultaneous role in optical degradation while enhancing electrical activation illustrates the complex nature of defect behavior in ion-implanted materials. This interplay underscores the importance of considering both optical and electronic properties when designing effective doping strategies.

The implications of these findings are twofold. On one hand, the generation of $Al_i$ interstitials during donor implantation increases the n-type conductivity, which is desirable for device applications that require efficient carrier injection. On the other hand, the possible formation of the associated $V_{III}$–n•donor complexes introduce optical losses and carrier scattering that may limit the performance of optoelectronic devices, such as deep-ultraviolet light-emitting diodes. This trade-off between conductivity, mobility, and optical quality highlights a fundamental challenge in the doping of wide-bandgap semiconductors: achieving high free-carrier concentrations while minimizing defect-induced nonradiative recombination and mobility reduction due to electron-defect scattering. Addressing this challenge requires a more detailed understanding of defect chemistry and diffusion, which our work helps to advance.

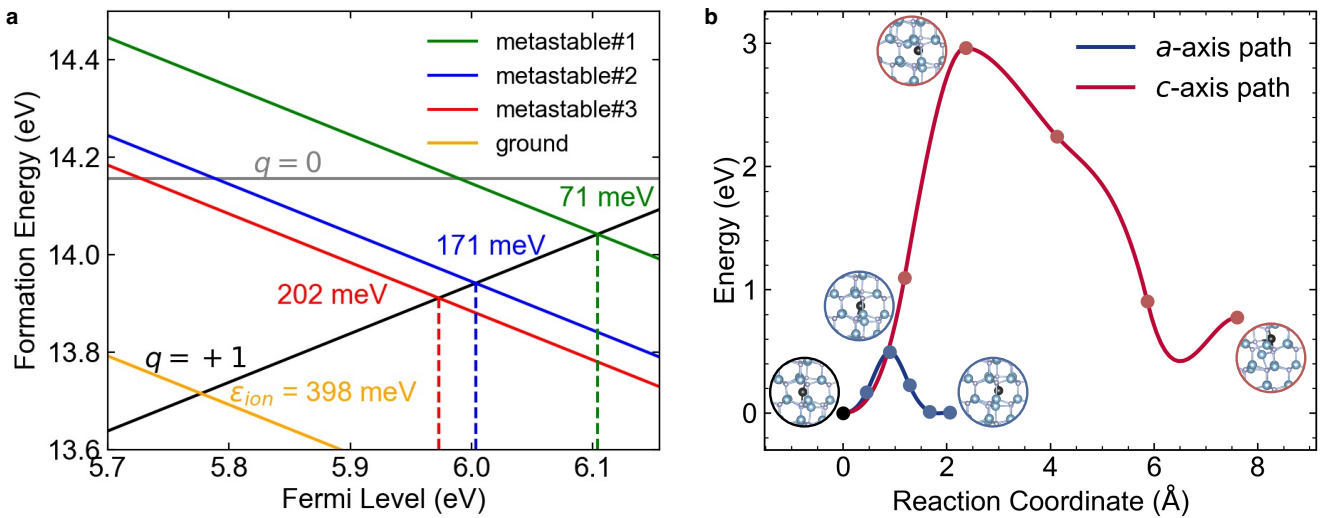

FIG. 4. (a) Formation energies of three metastable (#1, #2, and #3 in Fig. 3 (c)) and one stable states of $Al_i^{1-}$ with denoted ionization energies in corresponding colors. (b) cNEB calculations for the diffusion barriers of stable $Al_i^{1-}$ along $a$-axis (blue) and $c$-axis (red) directions, with the insets showing the atomic configurations along each path. The markedly higher barrier for the $c$-axis pathway demonstrates strong anisotropy of $Al_i$ mobility in AlN and its overall high diffusion barrier.

# Conclusion

In summary, our results provide a framework for rational doping strategies for possible shallow donors in AlN and related wide-bandgap materials. By understanding the interplay between defect complexes and interstitial donors, our findings suggest approaches to tune the balance between electrical and optical properties, either through controlled annealing treatments, co-doping schemes, or growth-condition optimizations that suppress the formation of detrimental complexes while preserving beneficial shallow donors. Future studies are necessary to explore the kinetic pathways for defect complex dissociation and the role of growth parameters—such as implantation energy, temperature, and chemical potential—in modulating the stability of these defects. This combined perspective of defect thermodynamics and kinetics is critical for advancing the design of high-performance electronic and optoelectronic devices based on AlN.

# Methods

We study defects and dopants in AlN with first-principles calculations based on density functional theory using the HSE hybrid exchange-correlation functional[63,64] and the Vienna *Ab initio* Simulation Package (VASP).[65] The mixing parameter is chosen as 0.33 to obtain good agreement with experiment both for the bandgap (6.275 eV, versus 6.2 eV in experiment[66]) and for the lattice constants ($a$ = 3.096 Å and $c$ = 4.955 Å, versus $a$ = 3.112 Å, $c$ = 4.982 Å in experiment[66]) of AlN. We choose the 128-atom supercell size given by the 3 × 3 × 2 multiple of the primitive cell. The plane-wave cutoff energy is set to 500 eV, and a Γ-centered 2 × 2 × 2 mesh is applied for the Brillouin zone sampling. Structural relaxations are controlled by the threshold of the residual forces within 0.01 eV/Å, and spin polarization is explicitly taken into account. For the structural and energetic analysis of defects in AlN, we use the ShakeNBreak (SNB) code[49,50] to generate 10 different starting structures by applying –40 % to 60 % bond distortions, as quantified by the dimensionless distortion factor $k$, up to the second nearest-neighbors of the defect atom. The distorted defect-neighbor distance ($d_n$) is defined as the product of the distortion factor ($k_n$) and the initial defect-neighbor distance ($d_0$): $d_n = k_n d_0$. The factor $k_n$ is calculated iteratively within a specific range $k_n = k_{min} + n\delta$, where $k\_min/k\_max$ is the minimum/maximum distortion factor, $\delta$ is the step-size increment parameter, and n as an integer stepper (0, 1, 2, ...) determined by the range $\frac{k_{max} - k_{min}}{\delta}$. We then randomly perturb all atoms in the supercell to allow for symmetry-breaking distortions, and perform structural relaxations with HSE to determine the relaxed stable and metastable defect structures. For the defect formation energies, we use the formula

$$E^f(D^q) = E^{tot}(D^q) - E^{tot}(AlN_{bulk}) + \sum n_i \mu_i + q(E_F + E_V) + \Delta E_{corr}(D^q), \quad (1)$$

where $E^{tot}(D^q)$ is the total energy of the system containing the defect and $E^{tot}(AlN_{bulk})$ is the energy of bulk AlN in the supercell size. The chemical potential term is given by $\sum n_i \mu_i$, where $n_i$ is the number of atoms that have been added/removed through the introduction of the defect, and $\mu_i$ is the chemical potential of the corresponding atom, referencing to the energy of the most stable elemental solid (e.g., $\mu_{Al}$) or gas (e.g., $\mu_N$) state. The Fermi level $E_F$ is aligned to the valence band maximum $E_V$, and $\Delta E_{corr}(D^q)$ is the finite-size correction term for charged defects.[67,68] The defect transition level $\epsilon(q/q')$ is the Fermi energy for which the formation energies of charge states $q$ and $q'$ are equal, and is identified by points at which the slope of the defect lines change in the formation energy versus Fermi level plot. The migration barrier between two different defect geometries is computed via the climbing images nudged elastic band (cNEB) method[54] as implemented in the VASP code using four or more intermediate images.

# ACKNOWLEDGMENTS

We acknowledge fruitful discussions with Drs. Ramon Collazo, Zlatko Sitar, and Pramod Reddy from NCSU. This work was supported by the Army Research Office under Grant No. W911NF-22-2-0176. Computational resources were provided by the National Energy Research Scientific Computing Center, which is supported by the Office of Science of the U.S. Department of Energy under Contract No. DE-AC02-05CH11231.